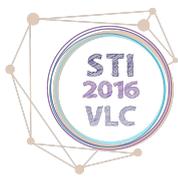

# Using network centrality measures to improve national journal classification lists[1]

Alesia Zuccala*, Nicolas Robinson-Garcia**, Rafael Repiso*** and Daniel Torres-Salinas****

* a.zuccala@hum.ku.dk
Royal School of Library and Information Science, University of Copenhagen, Birketinget 6, Copenhagen S, DK-2300 (Denmark)

** elrobin@ingenio.upv.es
INGENIO (CSIC-UPV), Universidad Politécnica de Valencia (Spain)

*** rafael.repiso@gmail.com
EC3metrics spin-off, Universidad de Granada (Spain)
Universidad Internacional de la Rioja (Spain)

**** torressalinas@gmail.com
EC3metrics spin-off, Universidad de Granada (Spain)

## INTRODUCTION

Research productivity for the scholar is often evaluated on the basis of his/her journal articles; however, specific journals are said to possess a higher measure of impact than others (Archambault & Larivière, 2009; Garfield, 2006; Glänzel & Moed, 2002). When a scholar decides where to publish, (s)he might consider a journal's impact factor. Although Garfield (1973) claimed that citation counts to individual articles will determine the impact factor of a journal, newer evidence points to the contrary: a journal with a high impact factor can also influence an article's readership and subsequent citation rates (Larivière & Gingras, 2010). This "chicken-and-egg" dispute (i.e., *which comes first – citations or impact?*) can be tested, but can still have negative consequences for how journals are selected, rated, listed, and used by policy-makers for developing measures of scholarly performance. For instance, in countries like Denmark and Spain classified journal lists are now being produced and used in the calculation of nationwide performance indicators. As a result, Danish and Spanish scholars are advised to contribute to journals of high "authority" (as in the former) or those within a high class (as in the latter). This can create a few problems.

First, a classification system that is designed to prize older, more established journals is problematic if it fails to acknowledge the role of the new journal. Scholarly research fields escalate and decline over time, and when a new area intensifies, sometimes a specialty journal is created. Data extracted from the Ulrich's periodical database for the period of 1900 to 1999 indicate "compound annual growth rates" for serials and has been used to suggest that "an increase of about 100 refereed papers per year world-wide, results in the launch of a new journal" (Mabe, 2001, p.159). Socio-political climates can further influence these growth rates, yet when a ranked list of journals is generated, the newer journal will inevitably start at a lower position. A scholar may then question or re-think his/her publication strategy. This type of decision making interferes unnecessarily with the natural flow of the learned society. According to Mabe (2001) the 'learned society' is essentially "a self-organizing information system that reflects the growth and specialization of knowledge" (p. 161). The role of a newly created journal is to function as a formal communication outlet where a gap is noticed within this natural, self-organizing system.

---

[1] Nicolas Robinson-Garcia is supported by a Juan de la Cierva-Formación Fellowship granted by the Spanish Ministry of Economy and Competitiveness.

The second problem rests with how journal lists are established and revised. Journals in emerging or peripheral fields might fail to make the list in the first place, while others are placed at a lower level or class. Here we will focus on two systems in particular: the 'authority list' related to the Danish bibliometric performance (BFI) system and the Integral Classification of Scientific Journals (known by its Spanish acronym, CIRC) for categorizing journals in Spain (Torres-Salinas, et al., 2010). The two systems differ because the first is based on peer-based judgements, while the latter is based on journal metrics and the presence of journal titles in international databases. The aim of this paper is to analyse the potential use of network centrality measures to identify possible mismatches of journal categories.

The paper is structured as follows. First, we describe the Danish and Spanish classification/rating systems. Second, we test a method for assessing and re-classifying journals in these systems, based on two complementary research techniques: 1) journal co-citation analysis and 2) social network centrality measures. These combined methods have previously been used to assess journals (e.g., Ni et al., 2011; Leydesdorff, 2007; McCain, 1991a, b), yet seem to be overlooked in this case as an informative policy-making tool.

**TWO APPROACHES: DANISH AUTHORITY VERSUS SPANISH METRICS**

*Danish Authority List:*
In 2009, Denmark developed an authority list of publications, and since this date, this list has been prepared and audited annually by over 350 researchers, across 68 'assigned' disciplines. Journals, book publishers and conference proceedings that pass the auditing process are categorized by the Danish academics as being either a 'level 1' outlet (normal) or a 'level 2' 'prestigious' outlet. According to the Danish bibliometric point system, known generally as the "BFI", publishing in a level 2 journal leads to a performance point of 3.0 while publishing in a level 1 journal earns a lower point of 1.0. The level 2 journal is expected to be that which covers a maximum of 20 % of world production of articles in the discipline to which it is assigned. Monographs and chaptered volumes also receive points, but we will not elaborate on these details, as they are not relevant to the scope of this study. The important aspect of the BFI system is that at the end of each year, cumulated points are used are used to determine how much of the Danish government's basic research funding (25% of the full allotment) is to be re-distributed amongst all universities (see Pedersen, 2010).

*Spanish CIRC Classification:*
In 2010, a group of Spanish bibliometric experts proposed a categorization of scientific journals for the Social Sciences and Humanities (Torres-Salinas et al., 2010). The classification aims at synthesizing the criteria of Spanish funding agencies for assessing journals from these areas and it is based on their inclusion and rank in a heterogeneous variety of tools and databases (Web of Science, Scopus, ERIH, etc.). Paradoxically, this classification has recently been included as a criterion in the Spanish performance-based evaluation system (Torres-Salinas & Repiso, 2016). CIRC classifies journals into five classes (A+, A, B, C and D). Journals are classified according to their compliance to certain criteria which are based on their inclusion in international databases and their Journal Impact Factor. It differentiates between Social Sciences and Humanities, as national evaluation standards also differ. Hence a journal may be categorized as B for Social Sciences and A for Humanities. Table 1 presents the CIRC classification criteria.

**Table 1**. Spanish CIRC criteria for classifying Social Sciences & Humanities journals.

|    | **Social Sciences** | **Humanities** |
|----|---------------------|----------------|
| **A+** | • Journals included in the first quartile in the JCR Social Sciences Edition according to their Impact Factor. | • Journals indexed in the A&HCI from Thomson Reuters and also positioned in the first quartile in Scopus according to their Impact per Publication (IPP) score.[2] |
| **A** | • Journals indexed in the SSCI or A&HCI, excluding those indexed in the fourth quartile of the JCR Social Sciences Edition according to their Impact Factor.<br>• Journals indexed in Scopus and positioned in the first quartile according to their IPP. | |
| **B** | • Journals included in the fourth quartile in the JCR Social Sciences Edition according to their Impact Factor. | • Humanities journals indexed in ERIH Plus (European Reference Index for Humanities). |
| | • Journals indexed in Scopus in the second, third and fourth quartile according to ther IPP score (excluding journals with IPP = 0).<br>• Spanish journals with a quality label recognized by the Spanish Foundation for Science and Technology (FECYT). | |
| **C** | • Journals indexed in Scopus with an IPP = 0.<br>• Social Sciences journals indexed in ERIH Plus.<br>• Journals indexed in the Regional Information System for Scientific Journals in Latin America, the Caribbean, Spain and Portugal catalogue (LATINDEX). | |
| **D** | • Journals included in the LATINDEX directory but not in its catalogue. | |

**METHODS**

Here we will compare the Danish and Spanish systems according to how each class of journal – i.e., level 1 and 2, or A+, A, B, C, D - 'fits' within a co-citation network. Our objective is to acquire information about the journal's network centrality within a specific field. The chosen field is Library and Information Science (LIS) and the method of data collection is as follows:

- A data sample (n=3,831 research articles) was extracted from all core indices of the Web of Science (WoS) for the publication year of 2015, and from the Subject Category: "Information Science and Library Science".
- The sample articles (n=3,831) were submitted to the VOSViewer mapping algorithm (Van Eck & Waltman, 2010) and used to produce a journal co-citation analysis based on a minimum citation threshold per journal set to 111 citations.

---

[2] Information obtained from http://www.journalmetrics.com

- A final co-citation network of 151 nodes was produced in VOSViewer (see Figure 1).
- A Pajek (*.net) file was then extracted from VOSViewer and used as input to the social network analysis and mapping tool, Netdraw (Borgatti, 2002).
- In Netdraw, a selection of node centrality measures, including *degree*, *betweenness*, *closeness*, and *eigenvector measures*, were calculated for each of the 151 nodes in the LIS journal co-citation network. Our research focuses mainly on the *eigenvector* and *betweenness* measures. *Eigenvector* centrality characterizes the global centrality of a node in a network and it is the most interesting indicator for our study, along with *betweenness* which indicates where a node possesses the shortest path between other node pairs, and shows the least correlation with the rest of the centrality indicators (Valente et al., 2008).

**Figure 1**. Journal co-citation network (n=151 nodes) from the WoS "Information and Library Science" category and each journal's Danish versus Spanish classifications.

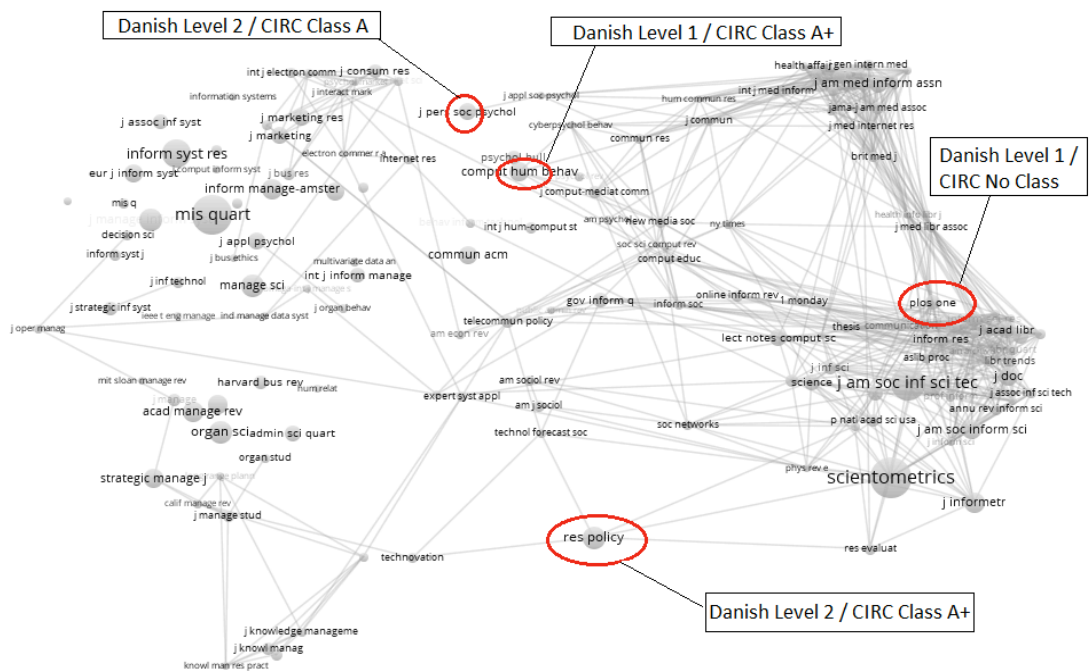

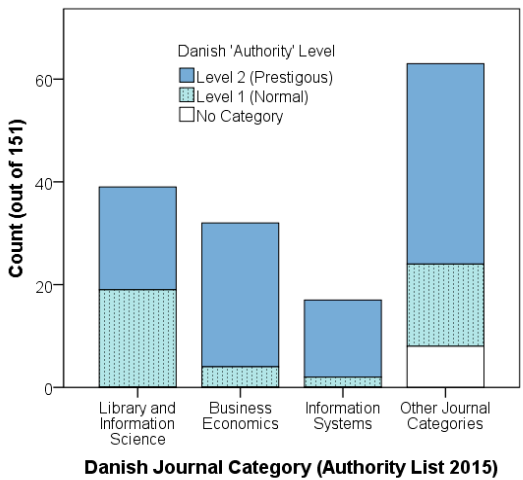

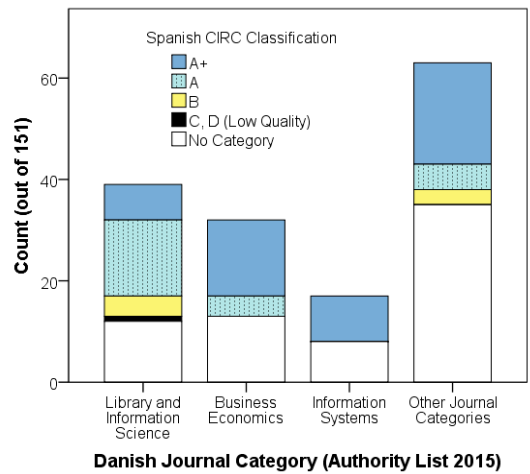

Figure 2                                   Figure 3

**RESULTS**

*Journal authority versus network centrality:*
Note from Figure 1 that we have captured the natural co-citation patterns of journal titles within "Information Science and Library Science.  Each journal in the co-citation network possesses either a central or peripheral role, or plays an 'in-between' role as a bridge between topics.   What we expect from a co-citation network of approximately 100-150 journals in a field is that new clusters will form over time; clusters that might lead to the creation of a new journal, or a central position for an existing one.

Figures 2, and 3, above indicate that the journal network has grown out of useful contributions from various fields. Here we present the fields according to the Danish journal categories, but compare all the actual journal classifications using the Danish Level 1 versus CIRC A+, A, B, C, D system.  58% of the co-cited journals in the network are level 2 journals from *LIS* (26%)*, business economics* (11%) and *information systems* (21%), and less than half of the journals are from other 'related' fields (e.g., *computer science, public health, science studies, media & communication, political science, medicine, psychology*).

**Figures 4 and 5**.  Eigenvector centrality values for journals in the 2015 LIS co-citation network: Danish "authority" levels versus Spanish CIRC classifications.

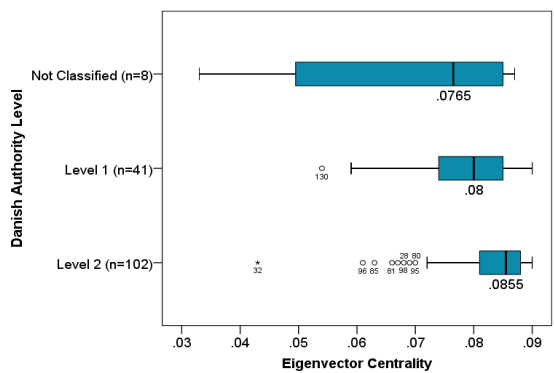
Figure 4

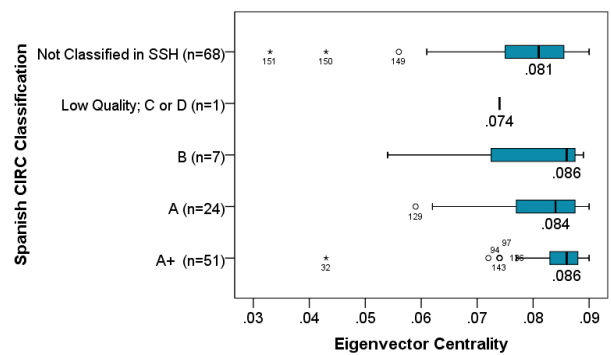
Figure 5.

The boxplots in Figures 4 and 5, show that the median *eigenvector* centrality values for journals classified by the Danish 'authority' system at level 1 or 2 differ slightly (.08 and .0855), while the A+, A and B median values in the Spanish system barely differ at all (.086, .084, and .086).  Note from Figure 4 that some journals in the third and fourth quartiles of the level 1 boxplot have *eigenvector* values that are just as high as those at or above the median value in the level 2 boxplot.  *ASLIB Proceedings* is one example of a journal that has an eigenvector of n=.09, which is higher than the level 2 boxplot median (.0855).  While it has been classified by the Danish system as a level 1 journal, it may have potential to be re-classified at some point to level 2.  One concern; however, is that it has had a name change to *ASLIB Journal of Information Management*; hence the consistency of its classification requires that the name change is accounted for in a new network analysis.

In Figure 5, the boxplot representing all journals with a B rating (Spanish CIRC) is skewed to the left. This indicates that more observations fall below the median, yet there are still a few B journals (above the comparative A .084 median) that play as much a central role in a network as an A or A+ journal For example, the journal *Information Research* is classified at level 2 in Denmark, but for the Spanish this is a B journal.

Journals classified in both the Danish level 2 and Spanish A+ categories with the highest *eigenvector* centrality measures (.09) include: *Journal of the American Society for Information Science and Technology (now Journal of the Association for Information Science and Technology)*, the *International Journal of Information Management* and *Scientometrics*. Journals classified in both the Danish level 1 and Spanish B categories with low eigenvector centrality measures include: *D-lib Magazine* (.06) and *Reference User Services Quarterly* (.05). The multiple outliers visible in the level 2 boxplot (see Figure 2) represent journals that play a less central role in library and information science, but have a 'prestigious' standing in other related fields (e.g., *Journal of Personality and Social Psychology*; *Journal of Marketing Research*).

**Figures 6 and 7**. Betweenness centrality value for journals in the 2015 LIS co-citation network: Danish "authority" levels versus Spanish CIRC classifications.

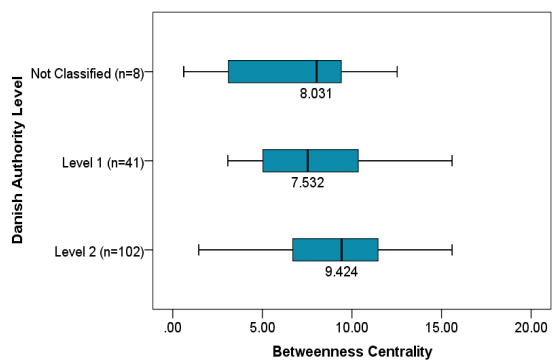
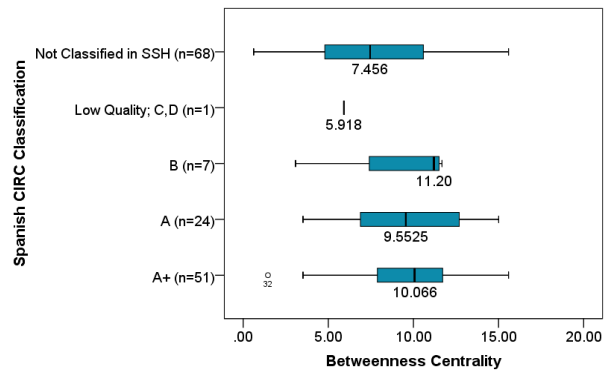

Figure 6 Figure 7

Figures 6 and 7 present boxplot distributions for the Danish and Spanish journal classes based on their *betweenness* measures in the network. In Figure 6, the boxplot for journals classified by the Danish system at level 1 is skewed to the right. This indicates that a higher measure of *betweenness* is observed more often for journals in this class than it is for 'prestigious' level 2 journals. Although many level 2 journals associated with outlier fields (e.g., psychology, medicine, economics) also have a high *betweenness* value. Note also from Figure 7 that the B journals classified by the Spanish system also tend to show a higher median value of *betweenness* than those from the A+ or A class.

In Table 2, below, the Danish and Spanish journal classification systems are compared again, and this time the level 1, 2, A+, A, B, C/D journals are distributed by quartiles according to their *eigenvector* centrality value. The results in table 2 may be examined in two different ways. For instance, we can focus on the journals that are considered 'prestigious' by both classification systems (level 2 in the Danish List, and $A^+$ or A in the Spanish CIRC classification). 39.2% of the level 2 journals in the Danish list are also included in the top 25% according to their *eigenvector*. A slightly higher value can be observed for the Spanish list (33.3%) and even higher if we only focus on the $A^+$ journals (49.0%). If we take an opposite view and examine the distribution of Q1 journals according to their *eigenvector* value, we observe that the highest share of these journals are categorised as prestigious (83.3% for the Danish class, 68.8% for the Spanish CIRC).

**Table 2**. Grouping of journals based on their Danish authority level and Spanish CIRC classification and their eigenquartile measures.

| Danish Authority List | | | | | |
|---|---|---|---|---|---|
| | Q1 | Q2 | Q3 | Q4 | Total |
| **2** | 40 | 22 | 27 | 13 | **102** |
| **1** | 7 | 4 | 14 | 16 | **41** |
| **0** | 1 | 2 | 1 | 4 | **8** |
| **Spanish CIRC Classification** | | | | | |
| $A^+$ | 25 | 11 | 9 | 6 | **51** |
| A | 8 | 4 | 7 | 5 | **24** |
| B | 2 | 2 | 1 | 2 | **7** |
| C/D | 0 | 0 | 0 | 1 | **1** |
| Not included* | 13 | 11 | 25 | 19 | **68** |
| **Total** | **48** | **28** | **42** | **33** | **151** |

\* These journals may not be included because a) they are not Social Sciences journals, or b) they have simply not been reported and are missing.

### *The 'evolving' journal*

In this part of our study we show how network centrality measures may be used as a support tool for re-classifying journals, particularly for those assigned to lists like the Danish and Spanish systems. Earlier we explained that the development of such lists can be problematic, because they might encourage scholars to publish in certain journals for the wrong reasons, or they can be too rigid if certain journals are kept a specific 'level' or class year after year. Our focal point for this analysis is the *Journal of Informetrics,* a relatively a young journal (featured in Figure 1), which was introduced in 2007, and has, within a short period of time, achieved a 'central' position in the field of LIS. This journal has been rated highly in both the 2015 versions of the Danish and Spanish journal classifications (i.e., level 2 and $A^+$ respectively).

With the *Journal of Informetrics* we have chosen to observe changes to its *eigenvector* centrality over time, alongside the *eigenvectors* of two more journals, *ASLIB Proceedings* and the *Annual Review of Information Science and Technology* (*ARIST*). To determine its evolution in the LIS field we have re-iterated similar journal co-citation networks for the earlier years of 2007 and 2011. The two new networks contain the top co-cited 151 nodes, like the 2015 map shown in Figure 1, and each was developed according to the same method as Figure 1. *ASLIB Proceedings* is currently categorized as a level 1 journal in the Danish authority list and A in the Spanish CIRC classification for social sciences. It is positioned in the second quartile according to its *eigenvector* value and serves as one example of a journal that could potentially be re-classified to level 2. ARIST, on the other hand is a level 2 journal that has been terminated as of 2011, but as of 2015 it was still classified as a level 1 (Danish) and $A^+$ journal (Spanish CIRC).

Figures 8 and 9 present the *eigenvector and betweenness* values for these three journals for the years 2007, 2011 and 2015. The *eigenvector* value shows the global centrality of a journal in a network, thus *ASLIB* but especially *ARIST* were core to the field in 2007. However, *Journal of Informetrics* has an incremental role. For all three journals, we see a convergence over time. The *eigenvector* value of ARIST decreases slightly, *ASLIB Proceedings* remains stable and the *Journal of Informetrics* increases.

**Figures 8 and 9**. *Eigenvector* and *betweenness* centrality measures of ARIST, ASLIB and Journal of Informetrics in three different periods within the LIS co-citation journal network

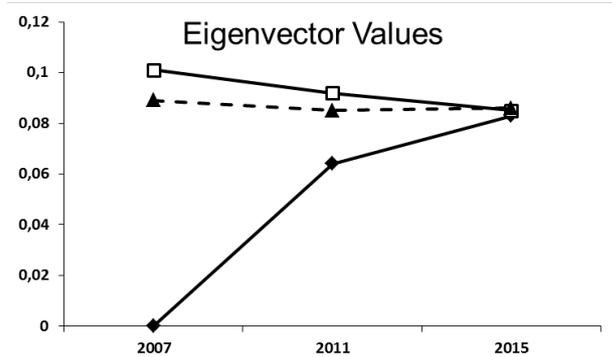
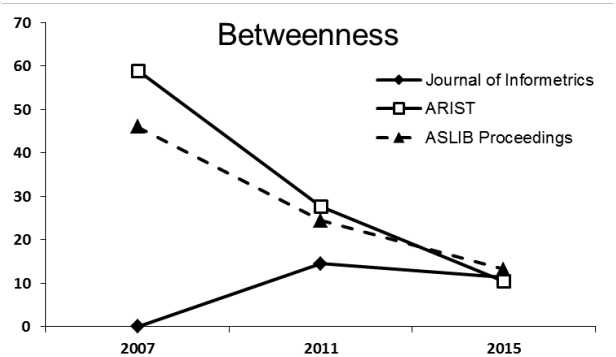

Figure 8.                                Figure 9.

In Figure 9, *betweenness* values show the local position of nodes in the network, and we are interested in this measure, because it can be used to identify journals that play a 'brokering' role (i.e., a link between topics). One might expect a new journal to play this role in its early stages, and indeed this is partially what happens with the *Journal of Informetrics* between 2007 and 2011. After 2011, its role as broker decreases slightly, but it is also within the same period (2011-215) when it achieves a higher *eigenvector* centrality, and becomes more central to LIS. An overall examination of each journal's changing *betweenness* measure shows that both *ARIST* and *ASLIB Proceedings* decrease in value, with ARIST showing the most dramatic decrease, while *Journal of Informetrics* increases slightly from 2007 to 2011, and then stabilizes.

**CONCLUSIONS AND FURTHER RESEARCH**

Thus far, both the Danish and Spanish national performance systems have relied either on traditional journal indicators, the presence of journals in international databases, or academic selection committees for developing journal classification lists. Notwithstanding problems associated with journal lists in the first place, this study shows that co-citation network centrality measures might be useful, particularly as a complementary policy tool. Here we conclude with a few policy-related recommendations and suggestions for further research.

While co-citation networks and their centrality measures are not sufficient for establishing the 'prestige' of a journal, they can still be used for making adjustments to a journal list. If a list is going to be reliable, it is likely to require periodic adjustments/revisions. The journal in question could be a new one, an older one, or one that has ceased to be active; however, a policy might be implemented whereby its centrality measure is observed across five-year periods within its natural co-citation network. This measure may determine whether or not the journal should be a) introduced at level 1, b) stay in its current level, c) be re-assigned as an 'evolved' level 2 journal, or d) be removed from a list entirely. In the case of a multidisciplinary journal, such as *Plos One* or *Nature*, the centrality measure will be different if it appears in different networks; thus consistently high centrality measures in multiple networks may be used to decide its classification.

Since we have focused on the *Journal of Informetrics*, it is important to note that the Danish 'authority' list was not established at the time this journal was first published in 2007. We do not know if it would have appeared in the Danish 'authority' list at level 1 in 2007 before it was 'promoted' to level 2. However, because it has experienced a rapid periphery-to-core transition within the LIS field, it seems to have earned its present level 2 class. This is not to say that other measures, factors, or dimensions contributed to its growth (see Haustein, 2007); however, its network eigenvector centrality is and may be used as a complementary 'objective' measure.

Last but not least, we need to consider future research. Overall there seems to be a general bias with older journals assigned to a higher class. For example, the founding year for journals listed as A+ in CIRC, is between the mid-1970s up to the mid-1980s, and as the establishing year of the journal gets higher (after 1990) the average class gets lower (as in the case of C journals for Humanities). The evolution of the *Journal of Informetrics* could be exceptional. Many new journals might not show a similar periphery-to-core evolution. It will be useful therefore to compare this journal's centrality shifts to other newer journals established at the same time, or to new journals in other fields.